\documentclass[10pt]{article}
\usepackage[table]{xcolor}
\usepackage{epsfig}
\usepackage{amssymb}
\usepackage{amsmath}
\usepackage{amsfonts}
\usepackage{url}
\usepackage{caption}
\usepackage{authblk}
\graphicspath{{Figures/}}
\usepackage[documents]{ragged2e}
\usepackage{booktabs} 
\usepackage[para]{footmisc}
\let\labelindent\relax

\usepackage[shortlabels]{enumitem}
\usepackage[]{algorithm2e}
\usepackage{fp}
\usepackage{multirow, array}
\SetAlgoSkip{}
\usepackage[pdfpagemode={UseOutlines},bookmarks=true,bookmarksopen=true,
bookmarksopenlevel=0,bookmarksnumbered=true,hypertexnames=false,
colorlinks,linkcolor={blue},citecolor={black},urlcolor={red},
pdfstartview={FitV},unicode,breaklinks=true]{hyperref}
\hypersetup{urlcolor=blue, colorlinks=true, breaklinks=true} 

\newcommand*{\TitleFont}{%
	\usefont{\encodingdefault}{\rmdefault}{b}{n}%
	\fontfamily{ptm}
	\fontsize{14}{16}%
	\selectfont}

\def\colorRed#1{   \FPeval{\grayshade}{100*#1}%
	\xdef\tempa{\grayshade}%
	\cellcolor{red!\tempa}\ifx #1>5\color{white}\fi{#1}}

\def\colorG#1{   \FPeval{\grayshade}{100*#1}%
	\xdef\tempa{\grayshade}%
	\cellcolor{green!\tempa}\ifx #1>5\color{white}\fi{#1}}

\author[1]{Dima Kagan\thanks{kagandi@post.bgu.ac.il}}
\author[1]{Yuval Elovici\thanks{elovici@post.bgu.ac.il}}
\author[2,3]{Michael Fire\thanks{fire@cs.washington.edu}}
\affil[1]{Department of Software and Information Systems Engineering, Ben-Gurion University of the Negev}
\affil[2]{Department of Computer Science \& Engineering, University of Washington}
\affil[3]{The eScience Institute, University of Washington}

\date{\today}
\begin{document}
	\justifying
	\newcommand{\s}[1]{{\color{red}{#1}}}

	\title{\TitleFont{Generic Anomalous Vertices Detection Utilizing a Link Prediction Algorithm}}

	\maketitle
	
	\begin{abstract}
		In the past decade, network structures have penetrated nearly every aspect of our lives.
		The detection of anomalous vertices in these networks has become increasingly important, such as in exposing computer network intruders or identifying fake online reviews.
		In this study, we present a novel unsupervised two-layered meta-classifier that can detect irregular vertices in complex networks solely by using features extracted from the network topology.
		Following the reasoning that a vertex with many improbable links has a higher likelihood of being anomalous,
		we employed our method on 10 networks of various scales,
		from a network of several dozen students to online social networks with millions of users.
		In every scenario, we were able to identify anomalous vertices with lower false positive rates and higher AUCs compared to other prevalent  methods.
		Moreover, we demonstrated that the presented algorithm is efficient both in revealing fake users and in disclosing the most influential people in social networks.
		
	\end{abstract}

	%
	%
	
	%
	%
	\providecommand{\plinks}[1]{\textbf{Project Links:} #1}

	\plinks{$\blacktriangleright$
		\href{https://youtu.be/SSRBaYw3gdM}{Video} $\blacktriangleright$
		\href{http://proj.ise.bgu.ac.il/sns/anomaly_detection.html}{Website} $\blacktriangleright$ \href{https://github.com/Kagandi/anomalous-vertices-detection/tree/master/data}{Data} $\blacktriangleright$ \href{https://github.com/Kagandi/anomalous-vertices-detection/}{Code}}

	\vspace{20pt}
	\providecommand{\keywords}[1]{\textbf{Keywords:} #1}
	
\section{Introduction}
Complex networks are defined as systems in nature and society whose structure is irregular, complex, and dynamically evolving in time.
The can consist of thousands, millions, or even billions of vertices and edges~\cite{albert2002statistical}.
These systems occur in every part of our daily life \cite{strogatz2001exploring, fire2016analyzing}, including electrical power grids, metabolic networks, food webs, the Internet,  co-authorship networks and online social networks \cite{strogatz2001exploring, newman2003structure}.
Analyzing the unique structures of such networks can be revealing; for example, an analysis of network structures can indicate how a virus will propagate most quickly in a computer network \cite{balthrop2004technological}, or which vertex malfunction in a power grid will affect more houses \cite{wang2003complex}.

Many studies have shown that vertices which deviate from normal behavior can offer important insights into a network \cite{akoglu2010oddball, bolton2002statistical,fire2012strangers, noble2003graph, papadimitriou2010web}.
For instance, Fire et al.~\cite{fire2012strangers} observed that fake profiles and bots in social networks have a higher probability than benign users of being connected to a greater number of communities.
Bolton et al.~\cite{bolton2002statistical} showed that e-commerce fraudsters behave differently from the expected norm.

\begin{figure*}		
	\begin{center}
		\includegraphics[width=\textwidth]{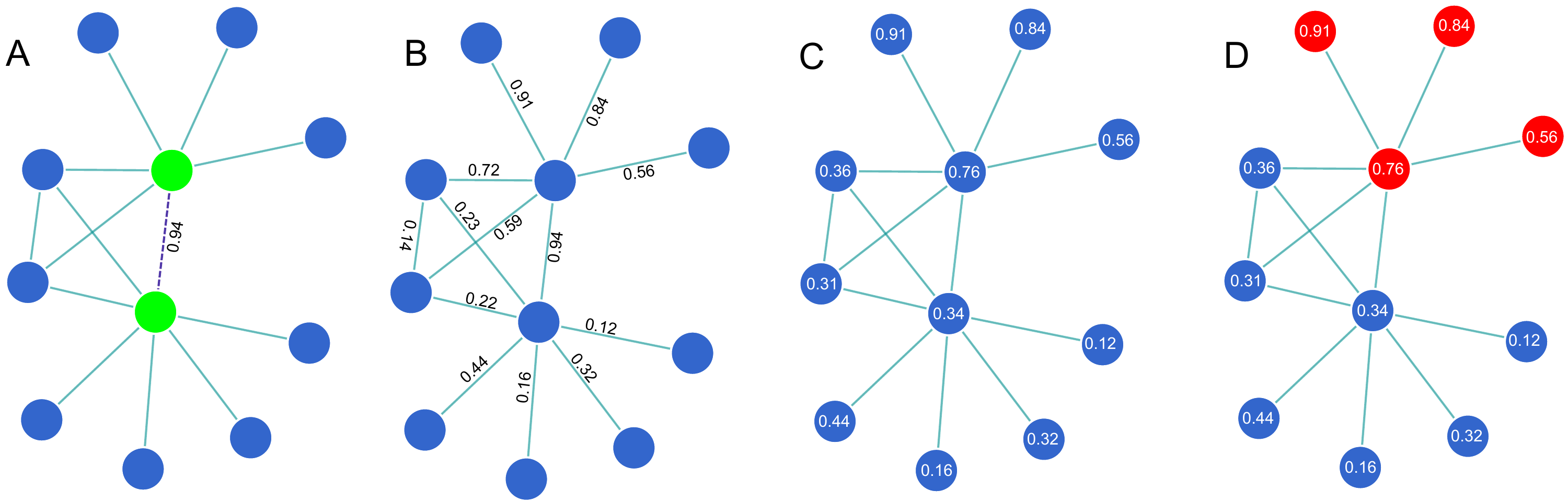}
	\end{center}

	\caption{Algorithm overview.
		(A) The link prediction classifier is trained to calculate the probability that an edge does not to exist in the graph. For example, the classifier can predict that the probability of an edge between the two green nodes not existing is 94\%. (B) We utilize the link prediction classifier to predict the probability of each edge not existing. (C)  We calculated the for each vertex the average probability that the vertex edges do not exist. (D) Vertices that have the highest average probability (the red vertices) are inspected.
	}\label{fig:method-flow} 

\end{figure*}

In this study, we introduce a novel generic unsupervised learning algorithm for the detection of anomalous vertices, utilizing a graph's topology (see Figure \ref{fig:method-flow}).
The algorithm consists of two main iterations.
In the first iteration, we create a link prediction classifier which is based only on the graph's topology.
This classifier can predict the probability of an edge existing in the network with high accuracy (see Section \ref{subsec:lp}).
In the second iteration, we generate a new set of meta-features based on the features created by the link prediction classifier (see Section \ref{subsec:anomaly}).
Then, we utilize these meta-features and construct an anomaly detection classifier.
Intuitively, the algorithm is based on the assumption that a vertex having many edges with low probabilities of existing (improbable edges) has a higher likelihood of being anomalous.
As a case in point, the presented algorithm was able to detect 89\% of the outliers in a real world social graph (see Section \ref{sec:anom-results}).

We evaluated our anomaly detection algorithm on 10 complex networks of three types: fully simulated networks, real world networks with simulated anomalous vertices, and real world networks with labeled anomalous vertices (see Section \ref{subsec:anom-eval} and Table \ref{tab:datasets}).
Our study results show that the proposed algorithm can successfully detect malicious users in complex networks in general, and more specifically, in online social networks. 
The presented algorithm succeeded in detecting various anomalies with high accuracy with low false positive rates.
Moreover, we showed that this algorithm can find influential people in social structures, indicating it could be applicable as a generic anomaly detection algorithm in additional domains.

\setlength{\arrayrulewidth}{0.3mm}
\setlength{\tabcolsep}{3pt}
\renewcommand{\arraystretch}{1}	
\begin{table}[b]
	\centering
	\caption{Social Network Datasets}\label{tab:datasets}
	\begin{tabular}{ |p{1.5cm}|p{1.3cm}|p{1.4cm}|l|p{0.8cm}|p{1cm}| }
		\hline	
		Network  & Directed & Vertices  & Links      & Date & Labels \\ \hline
		Academia & Yes      & 200,169   & 1,389,063  & 2011 & No     \\
		ArXiv    & No       & 34,546    & 421,578    & 2003 & No     \\
		Class    & Yes      & 53        & 179        & 1881 & Yes    \\
		DBLP     & No       & 1,665,850 & 13,504,952 & 2016 & No     \\
		Flixster & No       & 672,827   & 1,099,003  & 2012 & No     \\
		Twitter  & Yes      & 5,384,160 & 16,011,443 & 2012 & Yes    \\
		Yelp     & No       & 249,443   & 3,563,818  & 2016 & No     \\ \hline
	\end{tabular}
\end{table} 

The principal contributions of this study are the following: 
\begin{enumerate}[noitemsep]
	\item We demonstrate our successful incorporation of a link prediction technique into an anomaly detection model which requires almost no prior knowledge of the graph (see Section \ref{subsec:classifier}).
	For instance, our algorithm could detect anomalies in semi-simulated networks with an average AUC of 0.99.
	\item We present seven new network features that are good predictors for anomaly detection (see Section \ref{subsec:anomaly}).
	\item We offer valuable resources for future research by making this  study reproducible; we have published all of its code and data online, including the real world datasets containing labeled fake profiles (see Section \ref{sec:avail}).
\end{enumerate} 	
The remainder of this paper is organized as follows:
In Section \ref{rw}, we present an overview of relevant studies.
In Sections \ref{sec:anom-method} and \ref{subsec:datasets}, we describe how we constructed our algorithm and the datasets we used to evaluate it, respectively. 
Section \ref{subsec:anom-eval} provides a description of the experiments we performed to evaluate the presented method.  
In Section \ref{sec:anom-results}, we present our results.

\section{Literature Overview} 
\label{rw}
The detection of anomalies is extremely useful, because the discovery of irregularities without any prior knowledge or help of an expert is a necessity in many domains \cite{akoglu2015graph}.

The analysis of graphical data has grown in popularity \cite{eberle2007anomaly} over the past two decades, and this has been accompanied by increasing quantities of research on anomaly detection in complex networks.
However, many insights remain to be discovered, particularly in the structure-based method subgenre of anomaly detection. 	
One of the first studies that combined complex networks and anomaly detection was conducted by Noble and Cook~\cite{noble2003graph} in 2003.
Noble and Cook's method was based on the concept that substructures reoccur in graphs, which means anomalies are substructures that occur infrequently.

In 2005, Sun and Qu~\cite{sun2005neighborhood} proposed a method for detecting abnormal nodes in bipartite
graphs.
They calculated normality scores~(based on the neighborhood relevance score), where nodes with a lower normality score had a higher likelihood of being anomalous.
In 2007, Eberlea and Holder~\cite{eberle2007anomaly} proposed a mean of detecting fraud by discovering modifications, insertions, and deletions in graphs.
Unlike Noble and Cook, Eberlea and Holder looked for substructures that while similar to normative substructures, are still different.

In 2010, Papadimitriou et al.~\cite{papadimitriou2010web} presented a method that can identify anomalies in a web graph
that may occur as a result of malfunctions.
They proposed identifying outliers by comparing the similarity scores of two consecutive graphs against some threshold.
In 2010, Akoglu et al.~\cite{akoglu2010oddball} proposed a feature-based method to spot strange nodes in weighted graphs.
In order to detect anomalies, they defined a score that measures ``distance to fitting line'', considering nodes with the highest scores as outliers.
In 2012, Fire et al.~\cite{fire2012strangers} proposed a method for detecting fake profiles in online social networks based on anomalies in a fake user's social structure.
They trained their classifier on simulation of a fake profile that randomly sends friendship requests to other users in the network.
Recently, Hooi et al. \cite{hooi2016fraudar} presented FRAUDAR, a method for detection of ``camouflaged" malicious accounts. 
FRAUDAR utilizes density-based metrics in order to detect malicious accounts in bipartite networks.

In this work, we rely on a link prediction algorithm which is central to our anomaly detection method.
Link prediction is defined as the discovery of hidden or future links in a given social network~\cite{liben}.
The link prediction problem was first introduced by Liben-Nowell and Kleinberg~\cite{liben} in 2003 when they studied co-authorship networks and tried to predict future collaborations between researchers.
They proved that future links can be predicted with reasonable accuracy from network topology alone.

Cukierski et al.~\cite{cukierski2011graph} proposed a method that is based on supervised machine learning and uses 94 different features.
They discovered that a Random Forest classifier performed best out of all of the supervised machine learning methods evaluated.

Fire et al.~\cite{fire2013computationally} analyzed which topological features are more computationally efficient.
They tested a total of 54 different features that were divided into five subsets, using ten different datasets of online social networks.
They discovered that a smaller subset of features can be used to obtain results with relatively high AUCs \cite{fire2013computationally}.
Later, Fire et al. demonstrated that in many cases the benefits of using a large number of features is insignificant, and that by only using computationally efficient features, it is possible to get highly accurate classifications \cite{fire2013computationally}.

\section{Methods}
\label{sec:anom-method}


Studies conducted in the past several years indicate that malicious users typically present different behavioral patterns than benign users \cite{fire2012strangers, boshmaf2011socialbot, cao2012aiding}.
Motivated by the difference between the observed behavioral patterns of fake profiles and benign users, we developed a method to generate examples for our link classifier.
We aggregated the results from the link classifier for each vertex and created an additional set of features (see Section \ref{subsec:anomaly}).
We then, used the second set of features to build a meta-classifier  that identifies anomalous vertices in the graph.

\subsection{Constructing a Link Prediction Classifier} \label{subsec:lp}
As described in Section \ref{rw}, in the last decade researchers have proposed various methods for accurately predicting links in graphs \cite{ liben,cukierski2011graph, fire2013computationally, al2006link, brin2012reprint}.
In this study, we constructed a topology-based link prediction classifier based on the works of Fire et al. \cite{fire2013computationally} and Cukierski et al.~\cite{cukierski2011graph}. 
We extracted 19 different features,\footnote{In a large dataset, computing dozens of features can last several hours or even several days; to avoid extremely long computations we used only computationally efficient features \cite{fire2013computationally}.} 16 of which are used for directed graphs (all of the edge-based features except for \textit{TransitiveFriends} and \textit{AdamicAdarIndex} are used for both directed and undirected graphs) and eight are used for undirected graphs.
Prior to describing how the features were used, we provide the following definitions.
Let $G := (V,E)$ be a graph where $V$ is a set of the graph's vertices and $E$ is the set of the graph's edges.
Let $v\in V$; then $\Gamma(v)$ is defined as the neighborhood of vertex $v$, while $\Gamma_{in}(v)$, $\Gamma_{out}(v)$, and $\Gamma_{bi}(v)$ are defined as the inbound, outbound, and bidirectional set of neighbors, respectively.
\subsubsection{Feature Extraction}
\label{subsec:edge-features}
\begin{itemize}[noitemsep]
	\item \textbf{Total Friends} is the number of distinct friends between two vertices $v$ and $u$. 
	\begin{equation*}
	TotalFriends(v,u):=|\Gamma(v) \cup \Gamma(u)|
	\end{equation*}
	\item \textbf{Common Friends} represents the number of common friends between two vertices $v$ and $u$. 
	\begin{equation*}
	CommonFriends(v,u):=|\Gamma(v) \cap \Gamma(u)|
	\end{equation*}
	For a directed graph we define three variations of Common Friends:	\\
	$CommonFriends_{in}(v,u):=|\Gamma_{in}(v) \cap \Gamma_{in}(u)|$, 
	$CommonFriends_{out}(v,u):=|\Gamma_{out}(v) \cap  \Gamma_{out}(u)|$,  and
	$CommonFriends_{bi}(v,u):=|\Gamma_{bi}(v) \cap \Gamma_{bi}(u)|$.
	\item \textbf{Jaccard's Coefficient} \cite{cukierski2011graph,fire2013computationally, liben}  measures similarity between two groups of items.
	\begin{equation*}
	JaccardsCoefficient(v,u):=\frac{|\Gamma(v) \cap \Gamma(u)|}{|\Gamma(v) \cup \Gamma(u)|}
	\end{equation*}
	\item \textbf{Preferential Attachment Score} \cite{liben} is based on the idea that the rich get richer in social networks. 
	\begin{equation*}
	PreferentialAttachment(v,u):=|\Gamma(v)| \cdot |\Gamma(u)|
	\end{equation*}
	\item \textbf{Transitive Friends}\label{eq:trans} for a directed graph $G$ calculates the number of transitive friends of vertices  $v$ and $u$.
	\begin{equation*}
	TransitiveFriends(v,u):=|\Gamma(v)_{out}| \cap |\Gamma_{in}(u)|
	\end{equation*}
	\item \textbf{Opposite Direction Friends} indicates whether reciprocal connections exist between two vertices.
	\begin{equation*}
	OppositeDirectionFriends(v,u) :=
	\begin{cases}
	1, & \text{if}\ (u,v)\in E \\
	0, & \text{otherwise}
	\end{cases}
	\end{equation*}
	\item \textbf{Adamic-Adar Index}\label{eq:adamic} \cite{liben} measures how strongly two vertices are related.
	\begin{equation*}
	AdamicAdarIndex := \sum_{w \in \Gamma(u) \cap \Gamma(v)} \frac{1}{\log |\Gamma(w)|}
	\end{equation*}
\end{itemize}

\label{subsec:knn-features}
The kNN weight\footnote{This not the standard kNN acronym, but a set of weight functions defined by Cukierski et al. \cite{cukierski2011graph}.} features are general neighborhood and similarity-based features \cite{cukierski2011graph}.
They are based on the principle that as the number of friends goes up, the value of each individual friend decreases.

\begin{itemize}
	\item \textbf{Directed kNN Weights} are defined by two notations.
	Let $v,u \in V$, where $v$ edges weight will be $w_{in}(v):=\frac{1}{\sqrt{1 + |\Gamma_{in}(v)|}}$ and $w_{out}(v):=\frac{1}{\sqrt{1 + |\Gamma_{out}(v)|}}$, inbound and outbound, respectively.
	The weight of the connection between $v$ and $u$ can be measured using eight combinations of these weights:
	(a) $kNNW_1(v,u) := w_{in}(v) + w_{in}(u)$; 
	(b) $kNNW_2(v,u) := w_{in}(v) + w_{out}(u)$; 
	(c) $kNNW_3(v,u) := w_{out}(v) + w_{in}(u)$; 
	(d) $kNNW_4(v,u) := w_{out}(v) + w_{out}(u)$; 
	(e) $kNNW_5(v,u) := w_{in}(v) \cdot w_{in}(u)$; 
	(f) $kNNW_6(v,u) := w_{in}(v) \cdot w_{out}(u)$; 
	(g) $kNNW_7(v,u) := w_{out}(v) \cdot w_{in}(u)$; 
	and (h) $kNNW_8(v,u) := w_{out}(v) \cdot w_{out}(u)$.
	
	\item \textbf{Undirected kNN Weights}
	are defined similarly, but only for the neighbors.
	Let $v,u \in V$, where the weight of $v$ edges' will be $w(v):=\frac{1}{\sqrt{1 + |\Gamma(v)|}}$ and the weight of the connection between $v$ and $u$ will be measured by two combinations: (a) $kNNW_9(v,u) := w(v) + w(u)$ and (b) $kNNW_{10}(v,u) := w(v) \cdot w(u)$.
\end{itemize}
\subsubsection{Classifier Construction} 
\label{subsec:classifier}
Similar to Fire et al. \cite{fire2013computationally}, we trained the link classifier on the same number of negative and positive examples which are existing edges and non-existing edges respectively.
The negative examples represent real users and were selected randomly (degree distribution) from all of the existing edges in the graph.
The positive examples were selected as non-existing edges between two random vertices which were sampled uniformly to represent the edges of a malicious user in social network. 
After obtaining a set of positive (non-existing edges) and negative (random edges) examples, for each entry we extracted all of the features described in Sections \ref{subsec:edge-features}.
Finally, we used the Random Forest algorithm to construct the link prediction algorithm for our training sets.
We chose the Random Forest algorithm because previous link prediction studies~\cite{cukierski2011graph, fire2013computationally} demonstrated that, in most cases, it performs better than other classification algorithms at predicting links.
\subsection{Detecting Anomalous Vertices} \label{subsec:anomaly}	
After constructing a link prediction classifier for each graph, we utilized the classifier to build an unsupervised anomaly detection algorithm.
We used the trained classifier to calculate (for all the edges of the inspected vertices) the classifier's confidence that an edge does not exist.
Using this metric we calculated seven features that we used to identify anomalies.
\setlength{\textfloatsep}{0pt}
\begin{algorithm}[h]
	\small
	\SetAlgoLined
	\LinesNumbered
	\SetKwData{NeighborsNumber}{NeighborsNumber}\SetKwData{RandomVertex}{RandomVertex}\SetKwData{SelectedEdges}{SelectedEdges}\SetKwData{N}{N}\SetKwData{TempEdges}{TempEdges}
	\SetKwFunction{Set}{Set}\SetKwFunction{SampleNodes}{SampleNodes}\SetKwFunction{Edges}{Edges}\SetKwFunction{Add}{Add}
	\SetKwInOut{Input}{input}\SetKwInOut{Output}{output}
	\Input{Graph $G$, Number of nodes to sample $N$, Node label $Label$, Minimal number of friends $MinFriends$}
	\Output{Edges of Selected Vertices}
	\BlankLine
	$\SelectedEdges \leftarrow \Set()$\;
	\While{$N > 0$}{	
		\RandomVertex$\leftarrow$ \SampleNodes{$G$,1}\;
		\If {$\RandomVertex = Label ~ and ~ |\Gamma(\RandomVertex)| > MinFriends$}
		{
			$\TempEdges \leftarrow \Set()$\;
			\ForEach{Node $u$ in $\Gamma(\RandomVertex)$}{	
				\If {$|\Gamma(u)| > MinFriends$}
				{
					$\TempEdges\leftarrow \TempEdges + (\RandomVertex, u)$\;
				}
			}
			\If {$|\TempEdges| > MinFriends$}
			{
				$\SelectedEdges\leftarrow \SelectedEdges + \TempEdges$\;
				$\N\leftarrow \N - 1$\;
			}
		}
	}
	\Return $\SelectedEdges$;
	\caption{Sampling vertices from a graph.}\label{alg2} 
	
\end{algorithm}
\subsubsection{Anomaly Detection Features} \label{subsec:anom-features}
Let $p(v,u)$ be the probability that an edge does not exist, where $v, u \in V$, $(v,u) \in E$, and $EP(v) := \lbrace p(v,u) | u \in \Gamma(v)~and~ u,v \in V \rbrace$ is the set of vertex $v$ edge probabilities not existing.
\begin{itemize}[noitemsep]
	\item \textbf{Abnormality Vertex Probability} is defined as the probability of a vertex $v$ to be anomalous, which is equal to the average probability of its edges not existing. 
	\begin{equation*}
	P(v) := \frac{1}{|\Gamma(v)|}\sum\nolimits_{u \in \Gamma(v)}p(v,u) 
	\end{equation*}
	\item \textbf{Edges Probability STDV} is the standard deviation of a set of vertex $v$ edges' probability not existing.
	\begin{equation*}
	EdgesProbabilitySTDV(v) :=
	\sigma(EP(V)) 
	\end{equation*}
	\item \textbf{Sum Edge Label} is the number of vertex $v$ edges' that were labeled as anomalous; in other words, this is the number of edges $v$ with a $p$ higher than a defined $threshold$, which in this work was set to 0.8. 
	The goal of this feature is to detect cases where vertices have many anomalous edges, most of which are only slightly above the $threshold$, resulting in a relatively low $P$.
	\begin{equation*}
	SumEdgeLabel(v) := \sum\nolimits_{u \in \Gamma(v)} EdgeLabel(v,u) 
	\end{equation*}
	where we define the function $EdgeLabel(v,u)$ as:
	\begin{equation*}
	EdgeLabel(v,u) :=
	\begin{cases}
	0, & \text{if\ $p(v,u) < threshold$} \\
	1, & \text{otherwise}
	\end{cases}
	\end{equation*}
	\item \textbf{Mean Predicted Link Label} is the percent of $v$ edges that are labeled as anomalous.
	\begin{equation*}
	\label{eq:36}
		MeanPredictedLinkLabel(v) :=
		\frac{1}{|\Gamma(v)|}\sum\nolimits_{u \in \Gamma(v)} EdgeLabel(v,u)
	\end{equation*}
	\item \textbf{Predicted Label STDV} is the standard deviation of the classification of $v$ edges.
	\begin{equation*}
		PredictedLabelSTDV(v) := 
		\sigma(\lbrace EdgeLabel(v,u) | u \in \Gamma(v), u,v \in V \rbrace)
	\end{equation*}	
	\item \textbf{Edges Probability Median} is the median of set of vertex $v$ edges' probability of not existing.
	The advantage of median over mean is that it is not as sensitive to unusually large or small values. 
	\begin{equation*}
	EdgesProbabilityMedian(v) := 
	median(EP(V)) 
	\end{equation*}
	\item \textbf{Edge Count} is the number of edges that vertex $v$ has.
	An extremely low value may indicate that the results for vertex $v$ are statistically insignificant. 
	\begin{equation*}
	EdgeCount(v) := |\Gamma(v)|
	\end{equation*}
\end{itemize} 

We used the described features to rank all of the vertices by the different features and manually examined the top and bottom vertices, which had the highest and lowest likelihood of being anomalous.
\subsubsection{Test Set Generation}
First, we created the test set by sampling the edges of random vertices from the graph.
The sampling process works as described in Algorithm \ref{alg2}:
The algorithm starts by uniformly selecting one random vertex, $RandomVertex$ (line 3).
Next, we check if it has more neighbors than the minimal amount required ($MinFriends$).
In addition, for labeled graphs we also check if $RandomVertex$ has the desired label, in order to ensure that the test set has positive examples (line 4).
Then, we select all of the $RandomVertex$ neighbors that also have more than $MinFriends$ neighbors.
This constraint is used to ensure that the neighbors of $RandomVertex$ also were crawled and that the link features to be extracted are meaningful (lines 6-9).
If $RandomVertex$ has more than $MinFriends$ neighbors that have more than $MinFriends$ neighbors, then the selected edges are added to the test set (lines 11-14).

The goal of these steps was to select only edges between nodes which were likely fully crawled and had neighbors in the graph 
(i.e, have more than $MinFriends$ neighbors, which we set at three, the minimal number of vertices where we have a majority of edges of one of the class).
Vertices that have a small number of neighbors are less relevant, since there is not enough information to determine their behavior \cite{fire2012strangers}.
The algorithm continued to run until it added $N$ vertices to the test set.
In our experiments, we executed Algorithm \ref{alg2} twice for each network, the first time to extract positive samples and the second time to extract negative samples.
In both cases, we set $MinFriends$ to three.
\subsubsection{Training Set Generation}
Later, we sampled the link prediction classifier training set that was described in Section \ref{subsec:lp}.
The edge sampling for the link classifier worked as follows: Let $test\mbox{-}vertices$ be a set of all of the vertices that were selected by Algorithm \ref{alg2}, and if $(v,u) \in E$ is an edge, then $(v,u)$ can be part of the link classifier training set, if and only if, $u,v \notin test\mbox{-}vertices$. 

\section{Social Network Datasets} \label{subsec:datasets}

\label{itm:academia} \textbf{Academia.edu}\footnote{\url{https://www.academia.edu}} is a social platform for academics to share and follow research. 

\label{itm:arxiv} \textbf{ArXiv}\footnote{\url{https://www.arxiv.com}} is an ePrint service used in fields such as physics and computer science. 
We used the ArXiv HEP-PH (high energy physics phenomenology) citation graph that was released as part of the 2003 KDD Cup.\footnote{\url{https://snap.stanford.edu/data/cit-HepPh.html}}

\label{itm:class} \textbf{Boys' Friendship (Class Of 1880/81)}\footnote{\url{https://github.com/gephi/gephi/wiki/Datasets}} is a dataset which contains the friendship network of a German school class from 1880-81.
The dataset itself was generated by observing students, interviewing pupils and parents, and analyzing school essays \cite{heidler2014relationship}.
The dataset contains 12 outliers out of 53 students, which Heidler et al. defined as students who did not fit perfectly into their predicted position within the network structure.
The data contains three types of outliers: ``repeaters,'' who were four students who often led the games; ``sweets giver,'' a student who bought his peers' friendship with candies; and a specific group of seven students who were psychologically or physically handicapped, or socio-economically deprived.

\label{itm:DBLP}\textbf{DBLP}\footnote{\url{https://www.DBLP.com}} is the online reference for bibliographic information on major computer science publications. 
We used the DBLP dataset to build a co-authorship network.\footnote{\url{http://DBLP.uni-trier.de/xml/DBLP.xml.gz}}

\label{itm:flixster}\textbf{Flixster}\footnote{\url{https://www.flixster.com}} is a social movie site which allows users to share movie reviews and discover new movies.

\label{itm:twitter}\textbf{Twitter}\footnote{\url{https://www.twitter.com}} is an undirected online social network where people publish short messages and updates.
Currently, Twitter has 310 million monthly active users.\footnote{\url{https://about.twitter.com/company}}
According to recent reports, Twitter has a bot infestation problem \cite{nakedsecuritytwitter, Whyca60:online}.
We used a dedicated API crawler to obtain our dataset in 2014.\footnote{We limited the crawler to crawling a maximum of 1,000 friends and followers for every profile (see Section \ref{sec:avail}).
	This limitation is due to the fact that Twitter accounts can have an unlimited number of friends and followers, which in some cases can reach several million.
}

\label{itm:yelp}\textbf{Yelp}\footnote{\url{https://www.yelp.com}} is a web platform to help people find local businesses.
In addition, Yelp provides various social capabilities.
In 2016, Yelp published a dataset containing a social network of its users.\footnote{\url{https://www.yelp.com/dataset_challenge}}

\section{Experimental Evaluation}
\label{subsec:anom-eval}
We evaluated our algorithm on ten networks that categorized into three types of datasets (fully synthetic dataset, real world dataset with injected anomalies, and real world dataset with labeled anomalies).
In addition, due to hardware limitations, for each dataset we sampled a test set that contained 900 random existing vertices and 100 anomalous vertices.
The test set maintained the same 1:10 ratio between anomalous and normal vertices.
To reduce the variance of the results, we ran the algorithm 10 times on each dataset.
The more evaluations we performed, the smaller the variance in the results.
We evaluated the algorithm on the average result of these experiments.
To measure the algorithm's performance, we used 10-fold cross-validation to measure the TPR, FPR, precision, and AUC for all of the evaluations.
In addition, we measured the algorithm's precision at $k$ (precision@k) for k := 10, 100, 200, and 500.
Finally, we compared our method's performance to the Stranger algorithm \cite{fire2012strangers} and to a random algorithm in detecting anomalies. 
We only made the בomparison on the real world dataset since the training process of the Stranger algorithm is correlated with the way we generate our synthetic data.

\subsection{Simulation of Anomalous Vertices} \label{subsec:vertex-simulation}
Currently, there is a very limited number of publicly available datasets with known anomalies, and manual labeling is a challenging task \cite{akoglu2015graph}.
To deal with these issues and evaluate the proposed anomaly detection algorithm on various types of networks, we used simulated anomalous vertices (see Algorithm \ref{alg1}) for different scenarios.
Similar to previous studies \cite{fire2012strangers, boshmaf2011socialbot, cao2012aiding}, we generated anomalous vertices by randomly connecting them to other vertices in the network as follows.
First, we inserted a new simulated vertex into the graph (line 2).
Next, we generated $NeighborsNumber$, the number of edges to be created for the simulated vertex (line 3).
Then, we sampled random $NeighborsNumber$ vertices from the graph (line 4). 
Afterwards, we connected the newly inserted vertex to the sampled random vertices (lines 5-7).
The number of anomalous vertices in each graph was set to 10\%, which represents an estimation of the percentage of fake vertices in an average social network \cite{nakedsecuritytwitter, 2kreport2014facebook}.
\begin{algorithm}
	\small
	\SetAlgoLined
	\LinesNumbered
	\SetKwData{NeighborsNumber}{NeighborsNumber}\SetKwData{RandomVertices}{RandomVertices}\SetKwData{V}{SimulatedVerticesNumber}
	\SetKwFunction{PowerlawDist}{PowerlawDist}\SetKwFunction{SampleVertices}{SampleVertices}\SetKwFunction{AddVertex}{AddVertex}\SetKwFunction{AddEdge}{AddEdge}\SetKwFunction{Random}{Random}
	\SetKwInOut{Input}{input}\SetKwInOut{Output}{output}
	\Input{Graph $G$, having simulated vertex number $N$}
	\Output{Graph $G$ with $N$ simulated vertices}
	\BlankLine
	
	\For{$i\leftarrow 1$ \KwTo $N$}{
		\V$\leftarrow$ \AddVertex{$G$,$i$,$Fake$}\;	
		\NeighborsNumber$\leftarrow$ \Random{$G$.DegreeDistribution})\;
		\RandomVertices$\leftarrow$ \SampleVertices{$G$,\NeighborsNumber}\;
		\ForEach{Vertex $u$ in \RandomVertices}{
			\AddEdge{$v$,$u$}\;
		}
	}
	\caption{Adding anomalous vertices to a graph.}		\label{alg1} 
\end{algorithm}

\subsection{Fully Simulated Network Evaluation } 
To generate a fully simulated complex network we used the Barab\'{a}si-Albert model (BA model)~\cite{barabasi1999emergence}, which is a minimal' model that can generate scale-free networks.
We believe it should give a good indication of the performance of the method on various types of complex networks.
The generated networks were constructed according to the number of vertices and the average number of edges of a real world network to make them as close as possible to actual networks.
First, we generated BA networks that were 90\% of the size of the real networks that are described in Section \ref{subsec:datasets}.
Generating complex networks using the BA model requires two parameters: the number of vertices to be generated and the number of edges to be created for each vertex.
We used the number of vertices and the average number of edges of the ArXiv, DBLP, and Yelp datasets (see Table \ref{tab:datasets}) to generate the simulated networks.
Afterwards, we inserted anomalous vertices for the remaining 10\% (see Algorithm \ref{alg1}).

\subsection{Semi-Simulated Network Evaluation } 
The second dataset type was a semi-simulated network, which is a real world network with injected simulated anomalous vertices (see Algorithm \ref{alg1}).
The evaluation was conducted on the datasets of ArXiv, DBLP, Flixster, and Yelp (see Table \ref{tab:datasets}) with inserted anomalies.

\subsection{Real World Network Evaluation } 
The third dataset type we tested our method on was a real world network with labeled anomalous vertices.
We evaluated the graphs of the Boys' Friendship (referreed to as Class) and Twitter datasets (see Table \ref{tab:datasets}).
The Twitter data, by default, did not have any labels.
To create labels that can be considered ground truth, we crawled all of the profiles (without the edges) in the Twitter dataset, approximately one year after the initial crawling took place.
Similar to Hooi et al. \cite{hooi2016fraudar}, we considered all of the accounts that Twitter operators decided to block as a ground truth.
The most common reasons for suspension are spam, the account being hacked or compromised, and abusive tweets or behavior.\footnote{\url{https://support.twitter.com/articles/15790}\label{foot-1}}
As a result, we labeled all of the accounts which were suspended as malicious, and we considered them to be anomalous vertices in the graph.
In addition, we filtered all of the verified accounts from the dataset.
A verified account is an account of public interest, primarily those of celebrities, politicians, etc.\textsuperscript{\ref{foot-1}}
These were filtered because most of their connections do not represent regular users, and many times they are managed by some kind of third party \cite{Thatsnot48:online}. 

In the Class dataset, we observed that nearly all of the psychologically or physically handicapped students, (all except one) did not have neighbors in the graph.
This left us with three groups of outliers: the four ``repeaters,'' the single ``sweets giver,'' and two pupils who were socio-economically deprived.
Due to the small scale of the dataset, running the anomaly detection algorithm with a small number of repetitions can result in high variance rates. 
Therefore, to reduce the variance we ran our method 100 times on the dataset and calculated the average of the features presented in Section \ref{subsec:anomaly}.
In addition, every execution we tested contained only 10 vertices.

\section{Results}
\label{sec:anom-results}
\begin{figure}[h]
	\centering	
	\begin{center}
		\includegraphics[width=1\linewidth ]{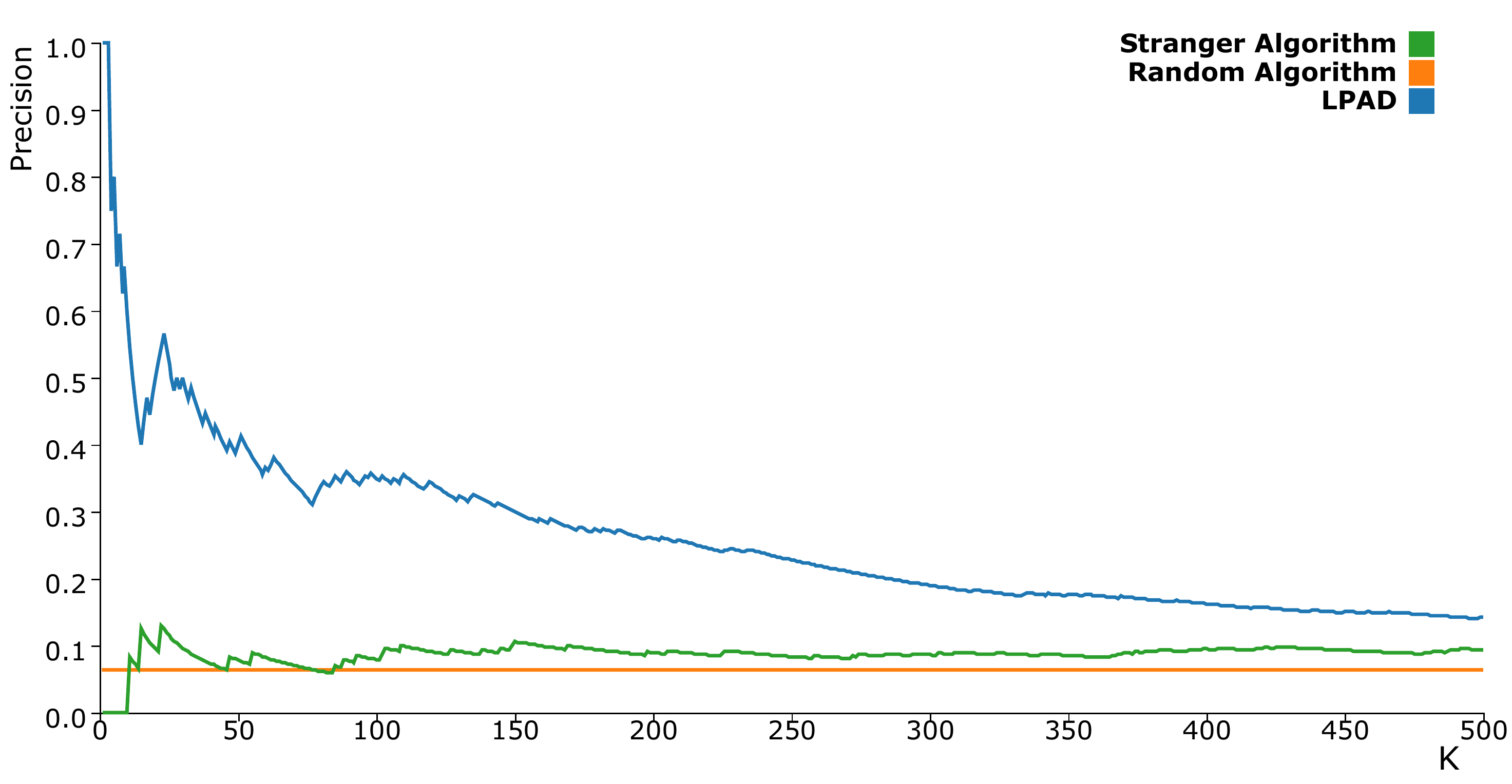}
		\caption{The blue, green and orange lines represent Twitter precision at K of LPAD, Stranger, and random algorithm respectively.} \label{fig:pk_twitter}
	\end{center}

\end{figure}

\begin{figure}[h]
	\begin{center}
		\includegraphics[width=\columnwidth]{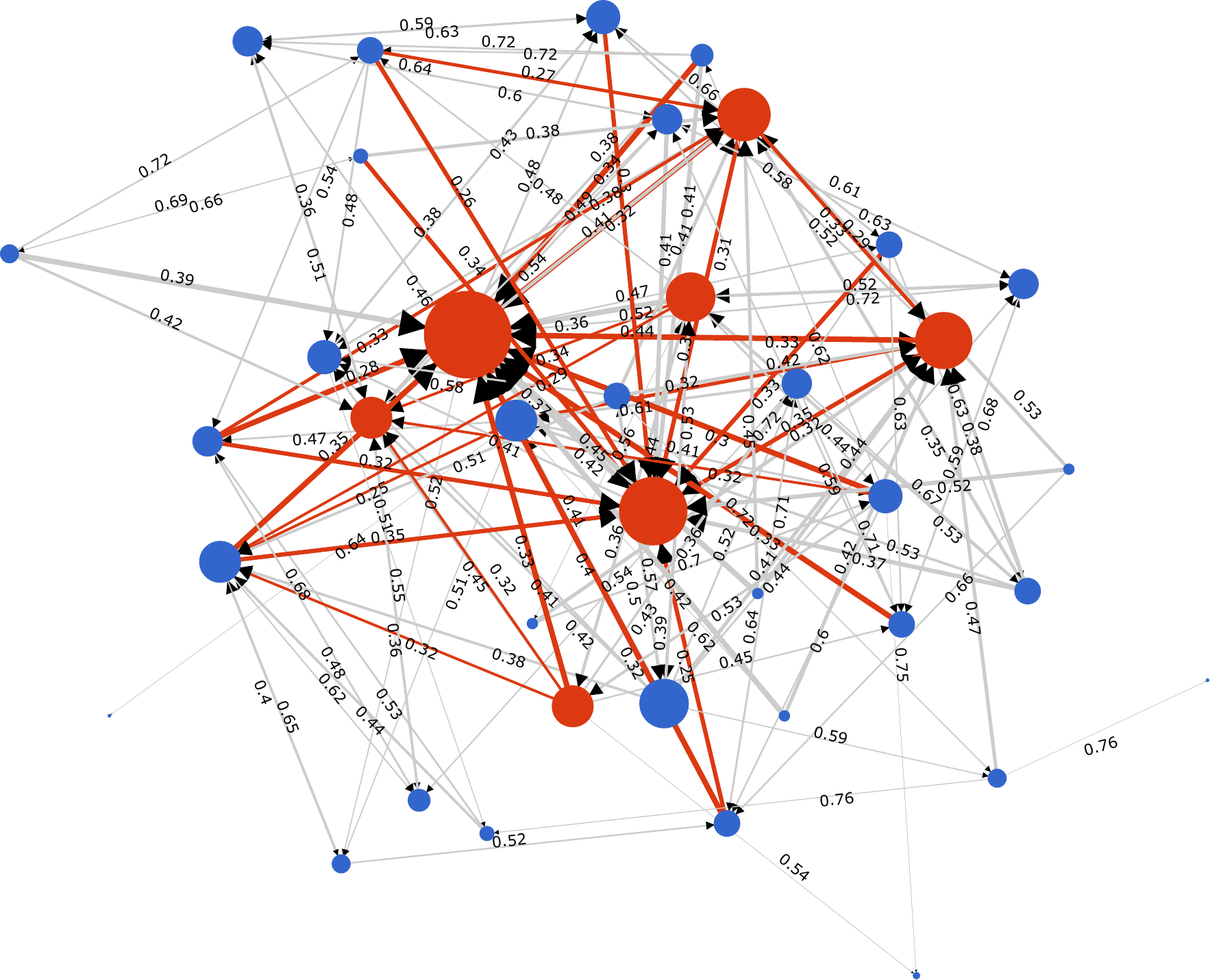}
	\end{center}
	\caption{Class network, where the red vertices represent the anomalous vertices (the boys who are the most central individuals in the friendship network), and the red edges are the edges that have the lowest probabilities of being fake. The graph demonstrates that almost all the red edges connected to the red vertices.}\label{fig:anom-graph}
\end{figure}

First, we evaluated the method on fully simulated networks with simulated anomalous vertices using a 10-fold cross-validation.
As can be seen in Table \ref{tab:sim}, for the three fully simulated networks, we obtained high AUCs and low FPRs.
Second, we evaluated the proposed method on semi-simulated graphs, i.e., real world networks with injected anomalous vertices.
We can see that the algorithm generated especially good results, with an average AUC of 0.99 and FPR of 0.021 (see Table \ref{tab:sim}).
\begin{table}[htb]
	\caption{Machine Learning Results Using Fully Simulated Networks and Semi-Simulated Networks with Simulated Anomalous Nodes}\label{tab:sim}
	\centering
	\begin{tabular}{|c|p{1.5cm}|p{1.0cm}|p{1.0cm}|p{1.5cm}|p{1.4cm}| }
		\hline &Network & AUC   & TPR   & FPR                 & Precision \\ \hline
		\multirow{3}{*}{\footnotesize{\rotatebox[origin=c]{40}{Simulation}}} &ArXiv                                                              & 0.991 & 0.889 & 0.011               & 0.904     \\
		&DBLP                                                               & 0.997 & 0.935 & 0.006               & 0.993     \\
		&Yelp                                                               & 0.993 & 0.917 & 0.007               & 0.937     \\ \hline
		\multirow{3}{*}{\footnotesize{\rotatebox[origin=c]{40}{Semi-Simulated }}}&Academia                                                           & 0.999 & 0.998 & $2.51 \cdot10^{-4}$ & 0.997     \\
		&Arxiv                                                              & 0.997 & 0.953 & 0.004               & 0.965     \\
		&DBLP                                                               & 0.997 & 0.940 & 0.005               & 0.995     \\
		&Flixster                                                           & 0.992 & 0.908 & 0.010               & 0.990     \\
		&Yelp                                                               & 0.996 & 0.941 & 0.005               & 0.958     \\ \hline
	\end{tabular}
\end{table} 
Third, we evaluated our algorithm on labeled real world data.
The first real world dataset was Twitter.
The results showing the classifier precision at $k$ average value are presented in Figure \ref{fig:pk_twitter}.
We can see that the precision at 10, 50, 100, 200, and 500 was 0.6, 0.4, 0.35, 0.26, and 0.142, respectively.
The second real world dataset was the Class network.
We found that six out of the seven students (precision@7=0.875) with the lowest $MeanPredictedLinkLabel$ were the ones that Heidler et al.~\cite{heidler2014relationship} referred to as the ``repeaters'' or the socio-economically deprived and defined as outliers (see Figure \ref{fig:anom-graph}).
Evaluating the algorithm using 10-fold cross-validation and the OneR algorithm, where the ``repeaters'' and socio-economically deprived students were labeled as a positive class, resulted in an AUC of 0.931, TPR of 0.91, and FPR of 0.15. In addition, we discovered that our method detects anomalies much better than Strangers \cite{fire2012strangers} in the Class dataset (see Table \ref{tab:comper}).
\begin{table}[b]
	\caption{Comparison of the current method (LPAD) with Strangers \cite{fire2012strangers} on the Class dataset (see Section \ref{subsec:datasets})}\label{tab:comper}
	\centering
	\begin{tabular}{|c|p{1.3cm}|p{1.3cm}|p{1.3cm}|p{1.3cm}|p{1.2cm}| }
		\hline Method & AUC   & TPR   & FPR   & Precision  \\ \hline
		LPAD\footnote{Link Prediction Anomaly Detection}              & 0.910  & 0.889 & 0.150  & 0.964  \\
		Strangers                                  & 0.714 & 0.439  & 0.006 &  1.000    \\ 
		ZeroR                                  & 0.279 & 0.000  & 0.000 & 0.000    \\ \hline
	\end{tabular}
\end{table}
\begin{table}[b]	
	\centering
	\caption{InfoGain Values for Different Features for Semi-Simulated Networks (darker color represents higher InfoGain) }\label{tab:rw-sim-ig}
	\resizebox{\columnwidth}{!}{%
		\begin{tabular}{|p{1.5cm}|p{1.9cm}|p{1.7cm}|p{1cm}|p{1.5cm}|p{1.7cm}|p{1.4cm}|p{0.9cm}| }
			\hline
			& Abnormality Vertex Probability & Probability Median & Sum Link Label & Mean Predicted Link Label & Probability STDV & Predicted Label STDV & Edge Count \\ 
			\hline     
			Academia     & \colorRed{0.47}                           & \colorRed{0.47}                & \colorRed{0.37}           & \colorRed{0.39}                      & \colorRed{0.29}             & \colorRed{0.1}                  & \colorRed{0}           \\
			Arxiv        & \colorRed{0.11}                           & \colorRed{0.1}                & \colorRed{0.1}            & \colorRed{0.11}                      & \colorRed{0.02}             & \colorRed{0.01}                 & \colorRed{0.01}        \\
			DBLP         & \colorRed{0.34}                           & \colorRed{0.23}              & \colorRed{0.33}           & \colorRed{0.32}                      & \colorRed{0.23}             & \colorRed{0.15}                 & \colorRed{0.04}        \\
			Flixster     & \colorRed{0.21}                           & \colorRed{0.19}               & \colorRed{0.2}            & \colorRed{0.21}                      & \colorRed{0.05}             & \colorRed{0.01}                & \colorRed{0.05}        \\
			Yelp         & \colorRed{0.18}                           & \colorRed{0.14}               & \colorRed{0.27}           & \colorRed{0.17}                      & \colorRed{0.25}             & \colorRed{0.06}                & \colorRed{0.06}        \\
			\hline
			Mean         & \colorG{0.34}                           & \colorG{0.31}               & \colorG{0.33}           & \colorG{0.32}                      & \colorG{0.2}              & \colorG{0.05}                 & \colorG{0.03}        \\
			STDV         & \colorG{0.18}                           & \colorG{0.19}               & \colorG{0.15}           & \colorG{0.16}                      & \colorG{0.12}             & \colorG{0.05}                 & 0\colorG{.02}       \\
			\hline
	\end{tabular}}
\end{table}
\begin{table}[h]
	\centering
	\caption{InfoGain Values for Different Features for Fully Simulated Networks (darker color represents higher InfoGain) \label{tab:full-sim-ig}}
	\resizebox{\columnwidth}{!}{%
		\begin{tabular}{|l|p{1.9cm}|p{1.7cm}|p{1.7cm}|p{1.6cm}|p{1.5cm}|p{1.3cm}|p{1.4cm}| }
			\hline   
			& Abnormality Vertex Probability & Probability STDV & Probability Median & Mean Predicted Link Label & Sum Link Label   & Edge Count      & Predicted Label STDV \\
			\hline   
			Arxiv        & \colorRed{0.18}                & \colorRed{0.216} & \colorRed{0.092}   & \colorRed{0}              & \colorRed{0}     & \colorRed{0.021} & \colorRed{0}         \\
			DBLP         & \colorRed{0.165}               & \colorRed{0.082} & \colorRed{0.146}   & \colorRed{0.148}          & \colorRed{0.124} & \colorRed{0.036} & \colorRed{0.072}     \\
			Yelp         & \colorRed{0.187}               & \colorRed{0.223} & \colorRed{0.105}   & \colorRed{0}              & \colorRed{0}     & \colorRed{0.027} & \colorRed{0}         \\
			\hline   
			Mean         & \colorG{0.18}                  & \colorG{0.17}    & \colorG{0.11}      & \colorG{0.05}             & \colorG{0.04}    & \colorG{0.03}    & \colorG{0.02}        \\
			STDV         & \colorG{0.01}                  & \colorG{0.08}    & \colorG{0.03}      & \colorG{0.09}             & \colorG{0.07}    & \colorG{0.01}    & \colorG{0.04}   \\ \hline       
	\end{tabular}}
\end{table}
To determine which of the new features we proposed in Section \ref{subsec:anomaly} have more influence, we analyzed their importance using Weka's information gain attribute selection algorithm.
From the results in Tables \ref{tab:rw-sim-ig} and Table \ref{tab:full-sim-ig} we can see that for both simulated and semi-simulated scenarios the most influential feature is $AbnormalityVertexProbability$.
\section{Discussion}\label{sec:discuss}
Upon analyzing the results presented in Section \ref{sec:anom-results}, we can conclude that the proposed anomaly detection algorithm fits well in the network security domain.
Our results demonstrate very low false positive rates, on average 0.006, in all of the tested scenarios.
In the security domain, a false positive is one of the most important metrics; online operators try to avoid false positives to ensure that legitimate users are not blocked.
Similarly, many social network operators prefer to sacrifice a true positive rather than have a relatively high false positive rate \cite{cao2012aiding}.

In the fully simulated network cases, we can see that the simulation results are correlated with the size of the networks.
As can be seen in Tables \ref{tab:datasets} and \ref{tab:sim}, we obtained better results for the larger datasets.
More specifically, the simulated network that was based on DBLP characteristics was the largest and had the best TPR, while the ArXiv-based simulation was the smallest dataset and had the lowest TPR.

In the Twitter case, we strongly believe there are substantial numbers of malicious accounts that Twitter operators have not discovered \cite{Whyca60:online}. 
These undiscovered malicious users translate into high false positives rates.
By manually sampling the false positives, we discovered that many of these profiles are inactive, and their tweets look like generated commercial content, whereas other profiles largely retweeted content from other users.
Because of the many unsuspended malicious accounts in Twitter~\cite{Whyca60:online}, we believe our method would perform better on a fully labeled dataset.
Yet even with these issues, we think Twitter is a good indicator of how well our method performs on real world data. According to the Twitter results (see Figure \ref{fig:pk_twitter}), we were able to detect fake Twitter profiles with precision at 100 of 35\%; this performance is considerably better than either Strangers or a random algorithm, which in this case resulted in approximately 8.0\% and 6.4\% precision, respectively.
One of the reasons we believe our method's performance in this case is much better than Stranger's is due to the data being incomplete. 
Stranger's algorithm is based mainly on community features, which are more sensitive to incomplete data than edge-based features.

The Class network results confirmed the previous research \cite{heidler2014relationship}.
The researchers described the ``repeaters'' as pupils who often led games and were strong, lively, and energetic, especially outside of the classroom.
They also mentioned that socio-economic status exhibited a strong influence on popularity.
In their work the researchers verified that the four ``repeaters'' and the ``sweets giver'' had a disproportionately high level of popularity. 
Our results show that the four ``repeaters'' had the strongest friendship ties of all the other pupils, which aligns with their findings.
The ``sweets giver,'' who also had high popularity, was just ranked in the middle, which also is reasonable, since some boys who looked like his friends only wanted candies, not friendship.

Lastly, according to the overall results, we can clearly state that our method can detect malicious profiles that act according to a random strategy.
We suspect, however, that the method would be less effective on malicious users that have specific targets and strategies.
We also found that it is more challenging to detect malicious users on networks like Twitter, where most of the users have some randomness in their behavior.
Such properties are more common in undirected networks where a user can follow anyone, without the need for the other side's consent.

Our results also indicate that the proposed method can be utilized outside of the security domain.
For instance, in a friendship graph, a vertex that has many edges with high probabilities of existing is a marker of a central person in the social group examined.
Moreover, we believe that the presented method could be used to detect hijacked profiles, when the hijacker starts to connect randomly to other vertices in the network.

\section{Conclusions}
\label{anom:conclusions}
The ability to detect anomalies has become increasingly important in understanding complex networks.
We present a novel generic method for detecting anomalous vertices based on features extracted from the network topology.
Our method combines cutting-edge techniques in link prediction, graph theory, and machine learning.

We evaluated our anomaly detection method on ten networks that can be categorized into three scenarios - fully simulated, semi-simulated, and real world networks
The evaluation results demonstrate that our anomaly detection model performed very well, particularly  in terms of AUC measure.
We demonstrated that in a real-life friendship graph, we can detect people who have the strongest friendship ties.
Moreover, we showed that our algorithm can be utilized to detect malicious users on Twitter.
We also showed that our method of detecting anomalous vertices outperforms other anomaly detection methods.
There is considerable potential in using the presented method for a wide range of applications, particularly 
in the cyber-security domain.
As for future directions, we plan to apply the algorithm to other types of networks, such as bipartite and weighted graphs.
Much can be revealed by investigating the behavior of anomalous vertices in complex networks.

\section{Availability}
\label{sec:avail}
This study is reproducible research. Therefore, the anonymous versions of the social network datasets and the study's code, including implementation, are available on the project's website\footnote{\url{http://proj.ise.bgu.ac.il/sns/anomaly_detection.html}} and repository.\footnote{\url{https://github.com/Kagandi/anomalous-vertices-detection}}

\section*{Acknowledgment}
\label{sec:ack}
We would like to thank Carol Teegarden and Robin Levy-Stevenson for editing and proofreading this article to completion. 
We also thank the Washington Research Foundation Fund for Innovation in Data-Intensive Discovery, and the Moore/Sloan Data Science Environment Project at the University of Washington for supporting this study.

	\bibliographystyle{abbrv}
	\bibliography{rp,privacyprotector,usenix}
\end{document}